# Modelling and Analysis of Car Following Algorithms for Fuel Economy Improvement in Connected and Autonomous Vehicles (CAVs)

Ozgenur Kavas-Torris[1], Levent Guvenc[1,2]

[1]Department of Mechanical and Aerospace Engineering, The Ohio State University, Automated Driving Lab (ADL), Columbus, OH 43212 USA
[2]Department of Electrical and Computer Engineering, The Ohio State University, Columbus, OH 43212 USA

Corresponding author: Ozgenur Kavas-Torris (e-mail: kavastorris.1@osu.edu)

## Abstract

Connectivity in ground vehicles allows vehicles to share crucial vehicle data, such as vehicle acceleration, with each other. Using sensors such as cameras, radars and lidars, on the other hand, the intravehicular distance between a leader vehicle and a host vehicle can be detected, as well as the relative speed. Cooperative Adaptive Cruise Control (CACC) builds upon ground vehicle connectivity and sensor information to form convoys with automated car following. CACC can also be used to improve fuel economy and mobility performance of vehicles in the said convoy. In this paper, 3 car following algorithms for fuel economy of CAVs are presented. An Adaptive Cruise Control (ACC) algorithm was designed as the benchmark model for comparison. A Cooperative Adaptive Cruise Control (CACC) was designed, which uses lead vehicle acceleration received through V2V in car following. an Ecological Cooperative Adaptive Cruise Control (Eco-CACC) model was developed that takes the erratic lead vehicle acceleration as a disturbance to be attenuated. A High Level (HL) controller was designed for decision making when the lead vehicle was an erratic driver. Model-in-the-Loop (MIL) and Hardware-in-the-Loop (HIL) simulations were run to test these car following algorithms for fuel economy performance. The results show that the HL controller was able to attain a smooth speed profile that consumed less fuel through using CACC and Eco-CACC than its ACC counterpart when the lead vehicle was erratic.

# 1. Introduction

Connectivity in ground vehicles allows vehicles to share crucial vehicle data, such as vehicle acceleration and speed, with each other. Using the on-board communication hardware that they have, connected vehicles (CVs) can share information with other CVs (Vehicle-to-Vehicle communication, V2V), as well as infrastructure (Vehicle-to-Infrastructure communication, V2I) and other traffic agents (Vehicle-to-Everything communication, V2X).

V2I models are well-studied and explored in the automotive industry. Signal Phase and Timing (SPaT) information from an upcoming traffic light can be used as part of the V2I technology for the longitudinal control of a Connected Vehicle (CV) to get smooth speed profiles and achieve fuel savings (Altan *et al.*, 2017), (Kavas-Torris *et al.*, 2021), (Cantas *et al.*, 2019a). Model Predictive Controllers (MPCs) can be utilized in conjunction with V2I technology to obtain both fuel savings and a shorter trip time (Asadi and Vahidi, 2011). Research has also been conducted on developing adaptive strategies in dealing with a dynamic roadway traffic environment while focusing on fuel savings (Wei *et al.*, 2019). Robust control and model regulation were also used for vehicle control (Aksun-Guvenc *et al.*, 2003), (Aksun-Güvenç and Guvenc, 2002), (Güvenç and Srinivasan, 1994). Parameter space with robustness was also utilized as another method for vehicle control (Güvenç *et al.*, 2017), (Aksun-Guvenc and Guvenc, 2001), (Aksun-Guvenc and Guvenc, 2002), (Emirler *et al.*, 2016), (Orun *et al.*, 2009), (Demirel and Güvenç, 2010), (Oncu *et al.*, 2007).

Other than connectivity, some vehicles also possess the feature of being an automated vehicle, called Connected and Automated vehicles (CAVs). CAVs are equipped with sensing information that provides information to them about their surroundings and make them aware of the objects in their vicinity. Front radars and cameras are two of the most popular and widely used sensors when it comes to detecting whether there is a vehicle in front of the ego vehicle or not. Radars and cameras can be used to measure the distance between a host/ego vehicle and a leader/preceding vehicle. This information can be utilized in Adaptive Cruise Control (ACC) application. ACC is an Advanced Driver Assistance System (ADAS), whose purpose is to keep either a constant set distance between the ego vehicle and the preceding vehicle or keep the ego vehicle speed the same as the preceding vehicle speed in order to avoid collisions (Aksun-Guvenc and Kural, 2006), (Kural and Güvenç, 2015), (Altay *et al.*, 2014), (Kural *et al.*, 2020), (Kural and Aksun Güvenç, 2010).

CAVs can have the best of both worlds by using both sensor information from their on-board sensors, as well as using information they receive from other CVs around them. In the academia, researchers have been studying how to effectively use both sensor and connectivity data together. Cooperative Adaptive Cruise Control (CACC) is one such method, where vehicle connectivity and sensor information are utilized by CAVs to form convoys to increase safety, improve fuel economy, decrease emissions. Detailed literature reviews have been conducted by researchers on previous research in the academia (Kianfar *et al.*, 2012) (Cantas *et al.*, 2019b). Additionally, CACC models are able to keep a smaller time gap between the vehicles in the convoy, which in turn can improve the roadway throughput (Liu *et al.*, 2020), (Wei *et al.*, 2019), (Cantas *et al.*, 2019b), (Ma *et al.*, 2020), (Ma *et al.*, 2021). Eco Cooperative Adaptive Cruise Control (Eco-CACC) is one method used for car following scenarios, where CAVs are driven in a cooperative manner by using V2V technology (Wang *et al.*, 2018) (Xiao and Gao, 2010) (Meier *et al.*, 2018) (Yang *et al.*, 2020).

In this paper, 3 different car following algorithms, namely ACC, CACC and Eco-CACC, were modeled to compare their performance when the lead vehicle was an erratic driver. A High Level (HL) controller was designed, which worked as a state flow diagram in Simulink and determined which car following model needed to be used at which instant during the simulation to ensure fuel savings.

The remainder of the paper is organized as follows: Methodology section explains the details about how the two different versions of the CACC model were developed, one being the Proportional-Derivate (PD) controller and the second one with a Model Predictive controller (MPC). Simulation Results section shows the performance of each CACC model for a given reference speed profile and quantifies how much fuel was consumed by the ego vehicle for both PD and MPC controlled CACC car following modes. Conclusions and Future Work section summarizes the work done, draws conclusions about the study and elaborates on how to extend this study for future work.

## 2. Methodology

For the analysis, an ACC, a CACC and an Eco-CACC model were designed. In the following sections, each car following is explained in detail.

### 2.1. Adaptive Cruise Control (ACC) Algorithm Design

Adaptive Cruise Control (ACC) models are widely used in car following applications. An ACC algorithm was designed, where the leader was being followed by the ego vehicle. Two vehicles following each other are shown in Figure 1. The dimensions of the ego and lead vehicles are neglected in the analysis for the sake of simplicity. $d_{st}$ is the standstill distance between these vehicles when their speeds drop to zero.

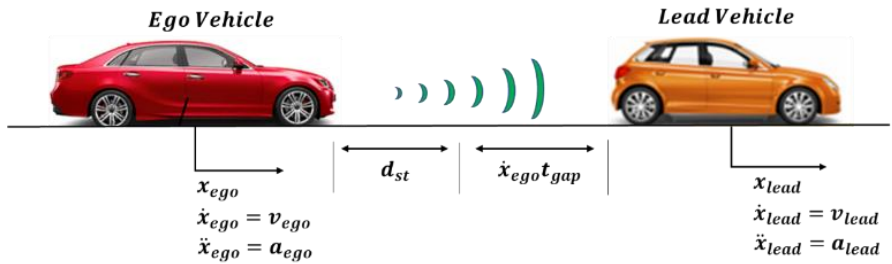

**Figure 1: Two vehicles following each other for ACC**

The desired car following distance (1) is,

$$x_{desired} = d_{st} + t_{gap}\dot{x}_{ego} \tag{1}$$

where $\dot{x}_{ego}$ is the ego vehicle speed and $t_{gap}$ is the time gap between the ego and lead vehicles. The actual distance between the two vehicles (2) is expressed as,

$$x_{actual} = x_{lead} - x_{ego} \tag{2}$$

The distance error, $e_x$ (3), can be defined as,

$$e_x = (x_{lead} - x_{ego}) - (d_{st} + t_{gap}\dot{x}_{ego}) \tag{3}$$

Taking the Laplace transform of Equation (3) with zero initial conditions and assuming that the standstill distance is zero results in (4),

$$\begin{aligned} E_x(s) &= X_{lead}(s) - (1 + st_{gap})X_{ego}(s) \\ E_x(s) &= X_{lead}(s) - H(s)X_{ego}(s) \end{aligned} \tag{4}$$

where the spacing policy $H(s)$ is given in Equation (5).

$$H(s) = (1 + st_{gap}) \tag{5}$$

The block diagram for the ACC algorithm with the Proportional-Derivative (PD) feedback controller $C_{fb}(s)$ and ego vehicle model $G(s)$ is given in Figure 2. In this model, the ego vehicle is equipped with a front looking radar and camera combination, which allows it to detect the lead vehicle as a target and determine its relative position and velocity.

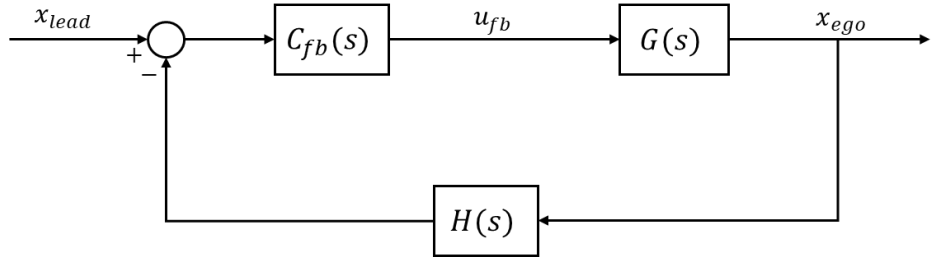

**Figure 2: ACC structure**

String stability is an important evaluation criterion for ACC and CACC systems. String stability of a string of vehicles refers to the property in which spacing error are guaranteed to attenuate as they propagate towards the tail of the string (Naus *et al.*, 2010), (Emirler *et al.*, 2018). It can be defined as follows,

$$\|SS_i(s)\|_\infty \leq 1 \Leftrightarrow \left|\frac{X_i(j\omega)}{X_{i-1}(j\omega)}\right| \leq 1, \forall \omega \tag{6}$$

where $SS_i(s)$ is the string stability transfer function for the $i^{th}$ vehicle in a string of vehicles and $X_i(j\omega)$ denotes the acceleration of the $i^{th}$ vehicle. The infinity norm of the string stability transfer function for each vehicle in the convoy needs to be smaller than or equal to 1 for string stability. The string stability transfer function for the ACC systems (7) can be written as follows,

$$SS_{ACC,i}(s) = \frac{C_{fb,i}(s)G_i(s)}{1 + C_{fb,i}(s)G_i(s)H_i(s)} \tag{7}$$

where $C_{fb,i}(s)$ is the feedback controller, $G_i(s)$ is the vehicle model and $H_i(s)$ is the spacing policy of the $i^{th}$ vehicle.

## 2.2. Cooperative Adaptive Cruise Control (CACC) Algorithm Design

For the Cooperative Adaptive Cruise Control (CACC) model, the leader vehicle was being followed by the ego vehicle and there was V2V communication between the vehicles. Two vehicles following each other for a CACC scenario is seen in **Figure 3**, where the lead vehicle acceleration was being received by the ego vehicle through V2V.

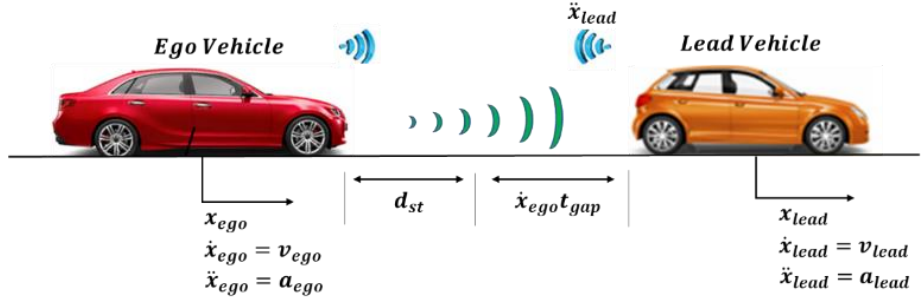

**Figure 3: Car following for CACC**

Similar to the ACC model, the feedback control for the CACC depends on the spacing policy, which is given in Equation (5). The block diagram for the CACC with the Proportional-Derivative (PD) feedback controller is given in **Figure 4**. In this model, the ego and the lead vehicle are both equipped with Dedicated Short Range Communication (DSRC) modems or a similar communication method that allows them to share information with each other. In this case, the lead vehicles shared its acceleration information with the ego vehicle. The ego vehicle is also equipped with a front looking radar and camera combination which allows it to detect the lead vehicle as a target and determine its relative position and velocity. The work of Emirler *et al.* (2018) was taken as a basis for the design of the CACC model presented in this section.

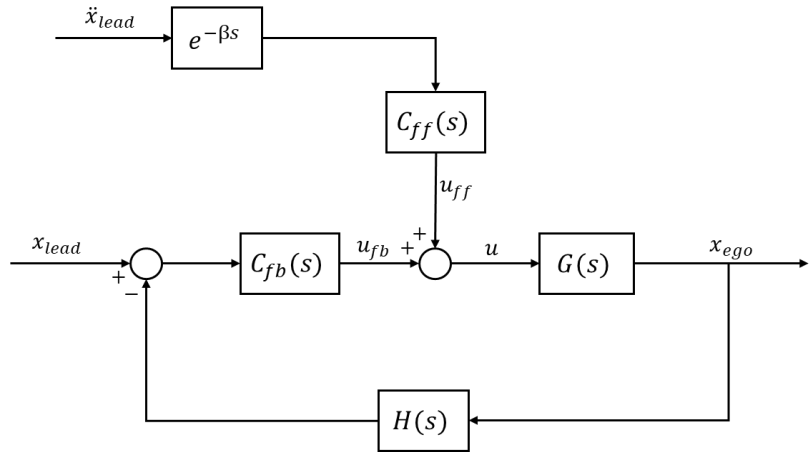

**Figure 4: CACC Structure**

The feedback controller called $C_{fb}(s)$ (8) is a PD controller and is expressed as follows,

$$C_{fb}(s) = K_p + K_d s \qquad (8)$$

where $K_p$ is the proportional gain and $K_d$ is the derivative gain. The acceleration from the lead vehicle was also used for the formulation, where $e^{-\beta s}$ represents the V2V communication delay between the lead and ego vehicle. The feedforward controller $C_{ff}(s)$ (9) is designed using the approach in Güvenç *et al.* (2012) and Naus *et al.* (2010), and is given by,

$$C_{ff}(s) = \frac{\tau s + 1}{t_{gap}s + 1} \quad (9)$$

where $1/\tau$ is the desired closed-loop bandwidth and $t_{gap}$ is the desired time gap (time headway). The vehicle model $G(s)$ (10) is given by,

$$G(s) = \frac{K_v}{s^2(\tau s + 1)} e^{-\kappa s} \quad (10)$$

where $K_v$ is the vehicle model gain, $\tau$ is the vehicle time constant and $e^{-\kappa s}$ represents the actuator delay in the ego vehicle.

Next, the string stability for this CACC model is analyzed, where the general string stability $SS_i(s)$ expression is given in (7). The string stability transfer function of this CACC system (11) is expressed as

$$SS_{CACC,i}(s) = \frac{\left(C_{fb,i}(s) + s^2 e^{-\beta s} C_{ff,i}(s)\right) G_i(s)}{1 + C_{fb,i}(s) G_i(s) H_i(s)} \quad (11)$$

The string stability of vehicles equipped with the CACC introduced here in a homogeneous convoy is found by plotting the string stability function as magnitude frequency response function. The vehicle parameters that are used in the calculation of the string stability function for 3 different cases are given in **Table *1***.

Table 1: Vehicle parameters for the CACC & ACC analysis

| CACC & ACC Parameters | Parameter Descriptions | Value Case 1 | Value Case 2 | Value Case 3 |
|---|---|---|---|---|
| $K_v$ | Vehicle model gain | 1 | 1 | 1 |
| $\tau$ | Vehicle time constant | 0.5 | 0.5 | 0.5 |
| $\kappa$ | Actuator delay [s] | 0.1 | 0 | 0 |
| $t_{gap}$ | Desired time gap [s] | 0.6 | 0.6 | 1 |
| $\beta$ | Communication delay [s] | 0.3 | 0.3 | 0.3 |

In Figure 5, the CACC and ACC infinity norms are plotted together to study the string stability. The results show the improvement in string stability offered by the CACC system over the ACC system.

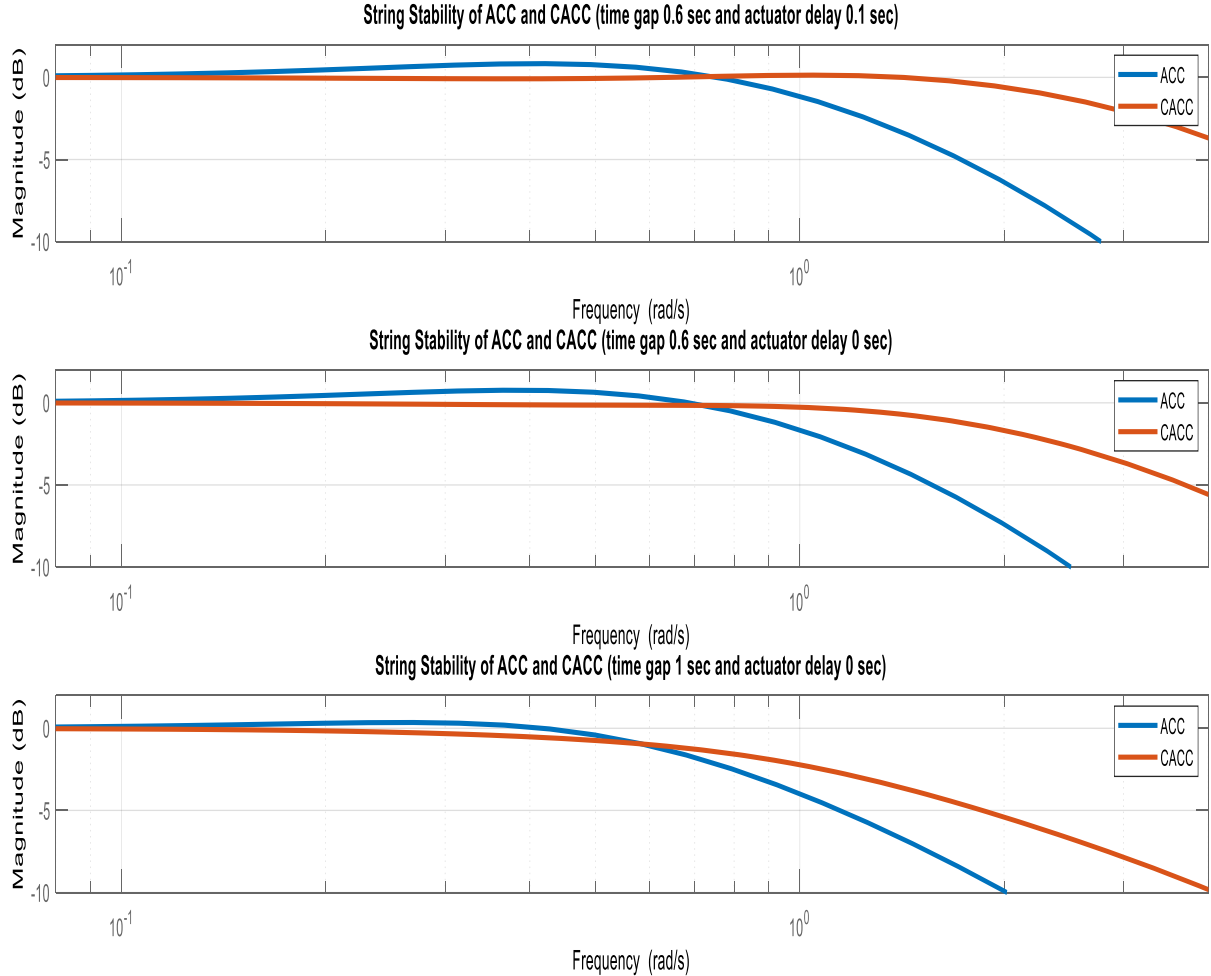

**Figure 5: String stability transfer function frequency response for ACC and CACC**

## 2.3. Ecological Cooperative Adaptive Cruise Control (Eco-CACC) Algorithm Design

When the lead vehicle follows an erratic speed profile, an Ecological Cooperative Adaptive Cruise Control (Eco-CACC) model is necessary. For this scenario, the preceding leader vehicle was being driven by an erratic driver, and the lead vehicle had unnecessary and frequent acceleration and decelerations. For this scenario, similar to the CACC algorithm, V2V was being utilized, where the acceleration of the lead vehicle was being received by the ego vehicle (**Figure *3***). However, unlike the CACC, the preceding vehicle acceleration was taken as a disturbance that was to be rejected for this model.

The block diagram for this Eco-CACC model when there is an erratic leader is in front of the ego vehicle can be seen in **Figure *6***. Similar to the CACC structure (**Figure *4***), the vehicle model is depicted as $G(s)$ (10). Similarly, the spacing policy is $H(s)$ (5), the feedback controller is $C_{fb}(s)$ (8) and the feedforward controller is $C_{ff}(s)$ (9).

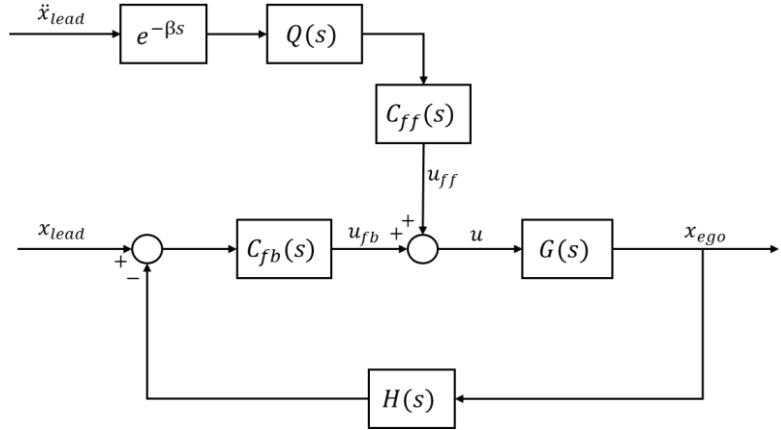

**Figure 6: CACC structure with erratic lead**

In this Eco-CACC structure, since the lead vehicle acceleration was taken as a disturbance, a filter was designed to attenuate the lead vehicle acceleration. The low pass filter with unity gain in this Eco-CACC structure is called $Q(s)$ (12) and is expressed as follows,

$$Q(s) = \frac{1}{f_c s + 1} \qquad (12)$$

where $f_c$ is the cutoff frequency. For the design of the filter $Q(s)$, as a first step, a simple speed profile for the leader vehicle was generated (Figure 7). On the left side, the leader accelerates to 15 m/s, travels at that speed for 400 m, then decelerates to a stop. On the right side, the leader accelerates to 15 m/s, then erratically speeds up and down in the form of a sine wave for 400 m, then decelerates to a stop.

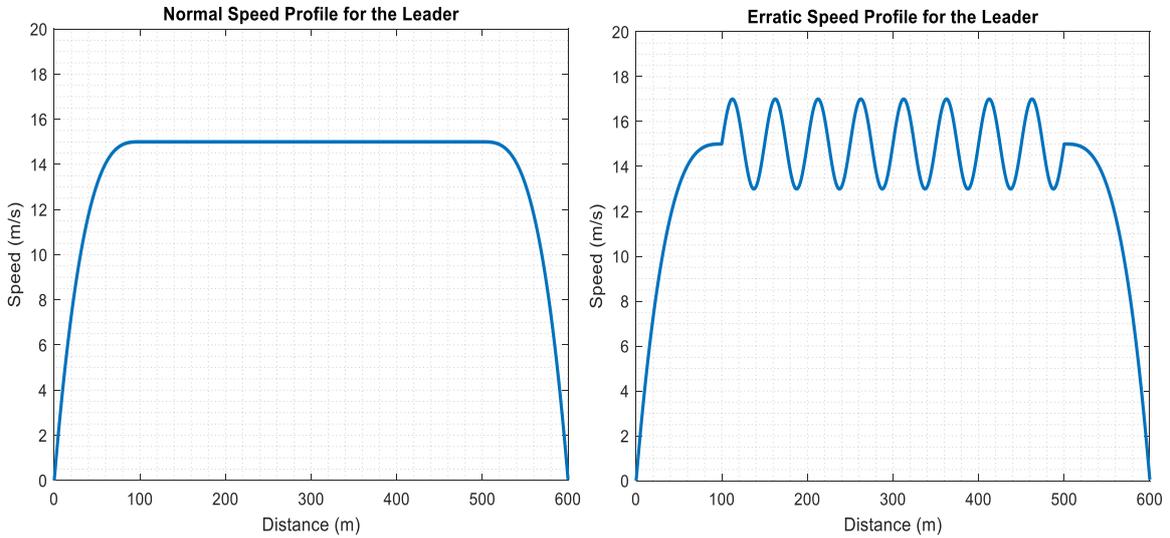

**Figure 7: Normal and erratic speed profile for the lead vehicle**

After these speed profiles seen in Figure 7 were generated, the Fast Fourier Transform (FFT) of both profiles were found and plotted together in Figure 8. The peak that differentiates the erratic profile from the normal speed profile is observed to be around 0.3 Hz. When the erratic leader follows the erratic speed profile given on the right side in **Figure 7**, a low pass filter $Q(s)$ should be designed such that the sine waves around 0.3 Hz are attenuated.

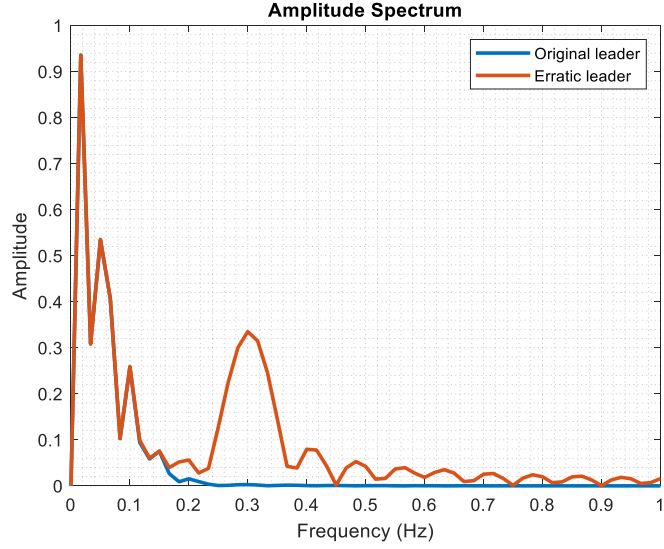

**Figure 8: FFT of the normal and erratic leader profile**

In order to show that filters with a bandwidth smaller than 0.3 Hz would be successful in attenuating the unwanted erratic lead vehicle acceleration, two different filters were designed. In the first filter called $Q_1(s)$, the bandwidth was kept small to attenuate the acceleration of the erratic lead vehicle. In the second filter called $Q_2(s)$, the bandwidth was kept large. The bode magnitude plot of the filters and where their cutoff frequencies are with respect to the sine wave frequency can be seen below in Figure 9. The bandwidth of filter $Q_1(s)$ was kept lower than the frequency of the sine wave to attenuate the sine wave.

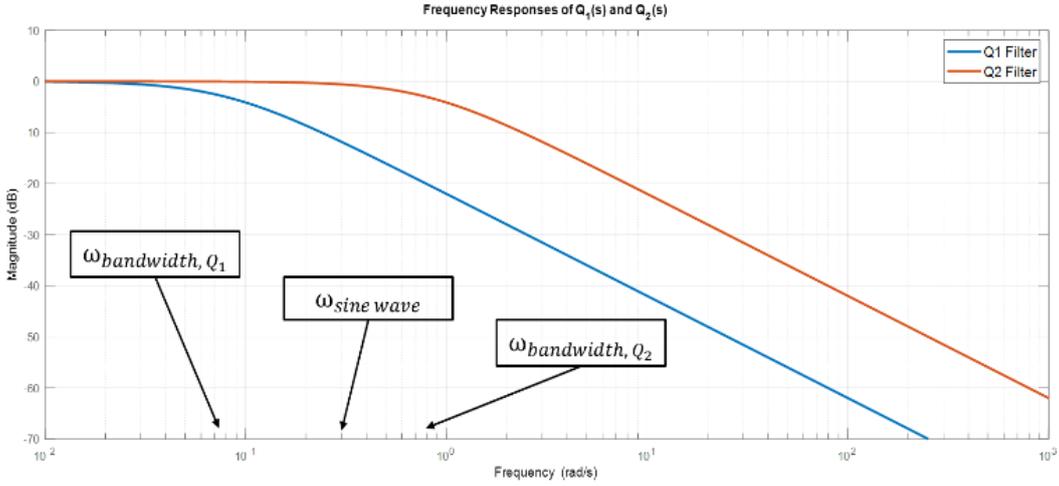

**Figure 9: Frequency responses of filters $Q_1(s)$ and $Q_2(s)$**

When it comes to the string stability of the Eco-CACC, the string stability transfer function of this Eco-CACC system is depicted as follows.

$$SS_{Eco-CACC,i}(s) = \frac{\left(C_{fb,i}(s) + s^2 e^{-\beta s} C_{ff,i}(s) Q(s)\right) G_i(s)}{1 + C_{fb,i}(s) G_i(s) H_i(s)} \qquad (13)$$

The string stability of vehicles equipped with Eco-CACC introduced in this section in a homogeneous convoy is found by plotting the string stability function as magnitude frequency response function, as seen in **Figure 10**. The Eco-CACC with the filter $Q(s)$, depicted in yellow, has infinity norm less than 1 for most frequencies with values going slightly above 1 for a very small frequency range.

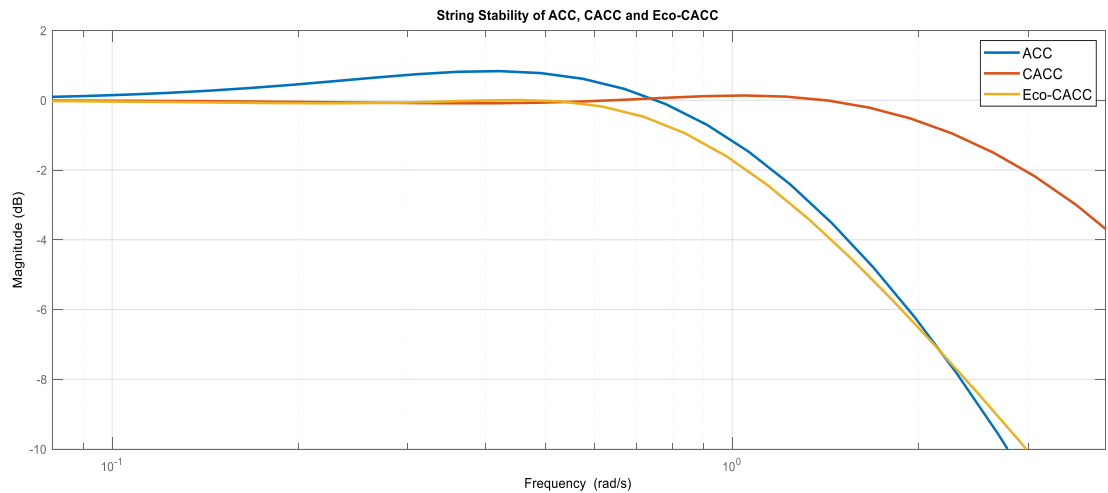

Figure 10: String stability frequency response for ACC, CACC and Eco-CACC

# 3. Simulation Results

Using the car following models presented in the Methodology section, firstly, Eco-CACC simulations were run with different $Q(s)$ filters. Additionally, Model-in-the-Loop (MIL) and Hardware-in-the-Loop (HIL) simulation results are also presented in this section.

## 3.1. Eco-CACC Simulations with Different $Q(s)$ Filters

A simulation scenario was modeled, where lead vehicle was following the normal speed profile shown on the left side in Figure 7 and the ego vehicle was following the lead vehicle. For the 1st simulation, the ego vehicle had only AC for car following. For the 2nd simulation, the ego vehicle used CACC with V2V for car following.

Looking at the results summarized in **Table 2**, it is seen that using the CACC model reduced the total fuel consumption of the ego vehicle by 8.22% compared to the ACC model. This result is expected, since CACC uses the lead vehicle acceleration as part of the calculation for the desired acceleration of the ego vehicle, whereas ACC only depends on the spacing error between the two vehicles.

Table 2: Fuel consumption comparison between ACC and CACC for a normal leader

| Car Following Model | Total Fuel Consumption [kg] | % Fuel Consumption Reduction wrt ACC |
|---|---|---|
| ACC | 0.0430 | - |
| CACC | 0.0394 | 8.22% |

Another simulation scenario was modeled, where lead vehicle had the erratic speed profile shown on the right side in Figure 7, and the ego vehicle was equipped with CACC, Eco-CACC with $Q_1(s)$ and Eco-CACC with $Q_2(s)$ filter. The fuel consumption of the ego vehicle for these 3 simulations are summarized in **Table 3**. CACC caused the ego vehicle to consume the most fuel when the leader was driving erratically. This result is expected, since when CACC was used, the ego vehicle tried to keep up with the erratic driving behavior of the lead vehicle rather than trying to attenuate the lead vehicle acceleration received through V2V. Eco-CACC performed better than CACC, as expected. Between the 2 filters, whose frequency response was given in Figure 9, Eco-CACC with smaller bandwidth filter $Q_1(s)$ consumed less fuel than Eco-CACC with $Q_2(s)$. Since Eco-CACC with $Q_1(s)$ performed better, $Q_1(s)$ was chosen as the filter $Q(s)$ for the Eco-CACC algorithm for the rest of the analysis.

Table 3: Fuel consumption comparison for car following models

| Car Following Model | Total Fuel Consumption [kg] | % Fuel Consumption Reduction wrt CACC |
|---|---|---|
| CACC | 0.1238 | - |
| Eco-CACC with $Q_1(s)$ | 0.0584 | 52.81% |
| Eco-CACC with $Q_2(s)$ | 0.0805 | 34.98% |

## 3.2. Model-in-the-Loop (MIL) Results

MIL simulations were run as co-simulations between Simulink and CarSim. For this simulation, information about the distance between the ego and lead vehicle, as well as the speed difference between the two vehicles, were gathered by using the soft camera and radar sensors of CarSim. The ego vehicle was equipped with a soft CarSim camera, and the distance between the two vehicles and the speed difference between the two vehicles were set as outputs of the CarSim model to be used for ACC, CACC and Eco-CACC applications.

For the first simulation, a realistic speed profile was given to the lead vehicle to follow. This speed profile started from 0 speed and included realistic accelerations and decelerations for the lead vehicle. Then, the ego vehicle was stationed behind the lead vehicle and followed the lead vehicle using the ACC model only.

Secondly, the ego vehicle was controlled by the High Level (HL) controller, where the ego vehicle was equipped with both CACC and Eco-CACC algorithms. The HL controller analyzed the driving behavior of the lead vehicle using the following equations

$$acc_{sum} = \sum_{i=1}^{n} acc_{lead}(t_i)^2 \qquad (14)$$

$$acc_{index} = \frac{acc_{sum}}{v_{lead}(t_n)} \qquad (15)$$

where $acc_{sum}$ is the summation of the squares of the lead vehicle acceleration in Equation (14). Acceleration index, $acc_{index}$, is found by dividing the $acc_{sum}$ by the lead vehicle speed at time $t_n$ (15). The HL controller starts by using CACC to follow the lead vehicle and saving information about the driving behavior of the lead vehicle. When the acceleration index goes above a certain value determined by prior simulations, then the lead vehicle is determined to be an erratic driver. When the lead vehicle is deemed erratic, then the HL controller activates the Eco-CACC algorithm to attenuate the erratic acceleration profile received from the lead vehicle. After following the erratic lead vehicle for a while, the ego vehicle checks to see if it can change lanes. Once the ego vehicle is able to change lanes, then it changes its lane. On the new lane, the ego vehicle is commanded to keep its speed constant.

The simulation results for the MIL simulation with V2V and traffic is seen in Figure 11. Before the lane change, the HL controller was able to follow the lead vehicle very closely by maintaining a smaller time gap and relative distance than the ACC only case. The desired time gap for the CACC model was set as 0.6 seconds and the desired time gap for the Eco-CACC model was set as 1 second. After the HL controller determined the ego vehicle could change its lane, the ego vehicle changed lanes around 97 seconds. Afterwards, the lead vehicle traveled at its erratic speed, whereas the ego vehicle kept its speed constant.

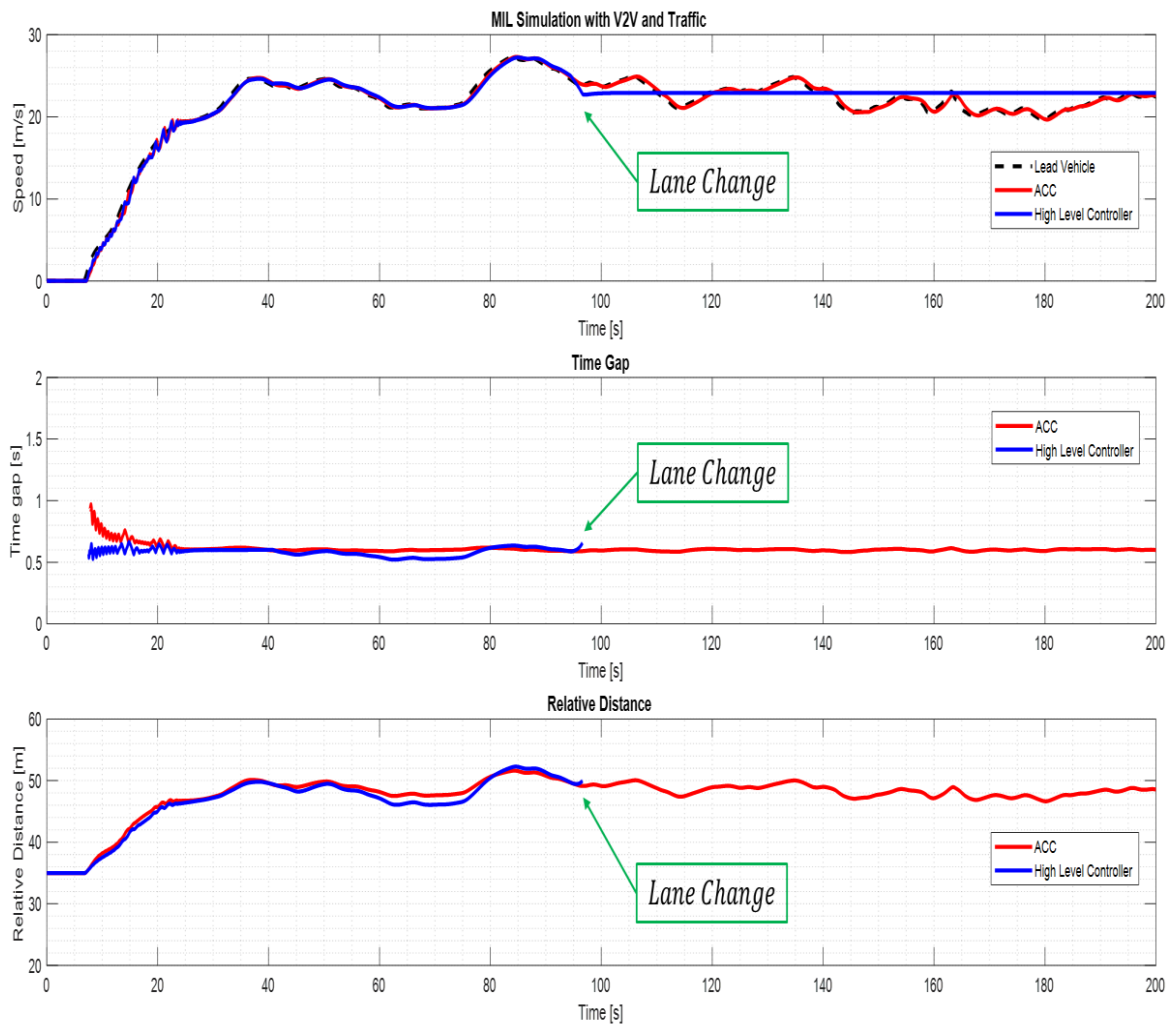

**Figure 11: MIL simulation with V2V and traffic results**

The results for these simulations are summarized in Table 4. Using the HL controller with CACC, Eco-CACC and Lane Change allowed the ego vehicle to have 3.99% fuel savings compared to the case where the ego vehicle only had ACC.

Table 4: Fuel consumption for the MIL simulation with V2V and traffic

| Model in MIL | Total Fuel Consumption [kg] | % Fuel Consumption Reduction wrt ACC |
|---|---|---|
| ACC | 0.135 | - |
| HL Controller | 0.129 | 3.99% |

## 3.3. Hardware-in-the-Loop (HIL) Results

Car following algorithms were then tested with real hardware in real time. The schematic of the HIL co-simulation environment for V2V testing can be seen in Figure 12. A time-dependent speed profile was prepared for the lead vehicle, and the acceleration of the lead vehicle was found by taking of the lead vehicle speed the derivative with respect to time. This lead vehicle acceleration was broadcasted by an Onboard Unit (OBU) modem with DSRC acting as the lead vehicle and was picked up by another OBU modem with DSRC acting as the ego vehicle. The lead vehicle acceleration was then sent to MicroAutoBox (MABX) that had the control algorithms running on it for the ego vehicle. Lead vehicle position and speed information were sent from SCALEXIO real time system acting as the ego vehicle to MABX. Then, using the information about the lead vehicle position, speed and acceleration, necessary control actions were calculated in MABX, and sent to both SCALEXIO and CarSim & Human Interface. Realistic vehicle information was then passed back to MABX to be used by the CACC and Eco-CACC algorithms to update the controls. Since hardware that is used in actual vehicles are utilized for the HIL setup with V2V, communication delays and actuator delays are more realistic and better aligned with real-life physical vehicle testing.

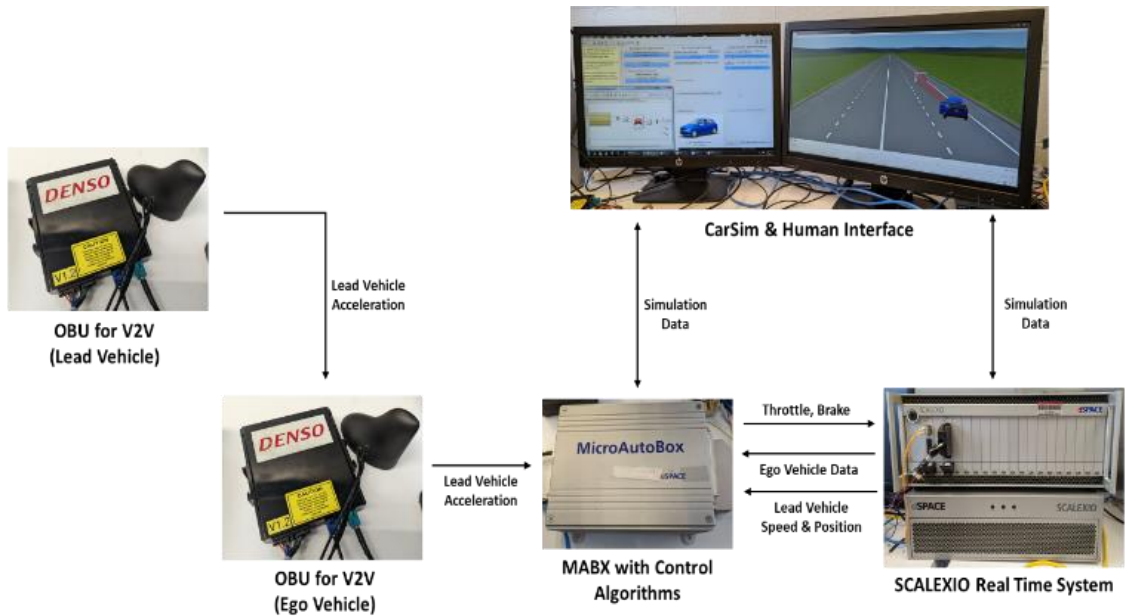

Figure 12: HIL setup for V2V testing

The results for the HIL tests for a CAV with V2V in traffic while following a lead vehicle can be seen in Figure 13. The lead vehicle speed profile is marked with a dashed line in black. When the ego vehicle was using only the ACC model, the ego vehicle kept following the lead vehicle and trying to keep a constant time gap of 0.6 seconds. When the HL controller was used, the ego vehicle first used CACC to follow the lead vehicle by trying to keep a time gap of 0.6 seconds. Additionally, the lead vehicle acceleration was sent to the ego vehicle through the OBU modems with DSRC communication, effectively creating realistic communication delays during the HIL simulation. When the lead vehicle acceleration index was found to be high, meaning that the lead vehicle was deemed to be erratic, then the Eco-CACC model was activated. When Eco-CACC was active, lead vehicle acceleration that was received through DSRC communication using OBU modems was used by the Eco-CACC and the time gap between the vehicles was kept around 1 second. Once the ego vehicle was able to change its lane, the ego vehicle changed lanes and passed the lead vehicle and kept its speed constant.

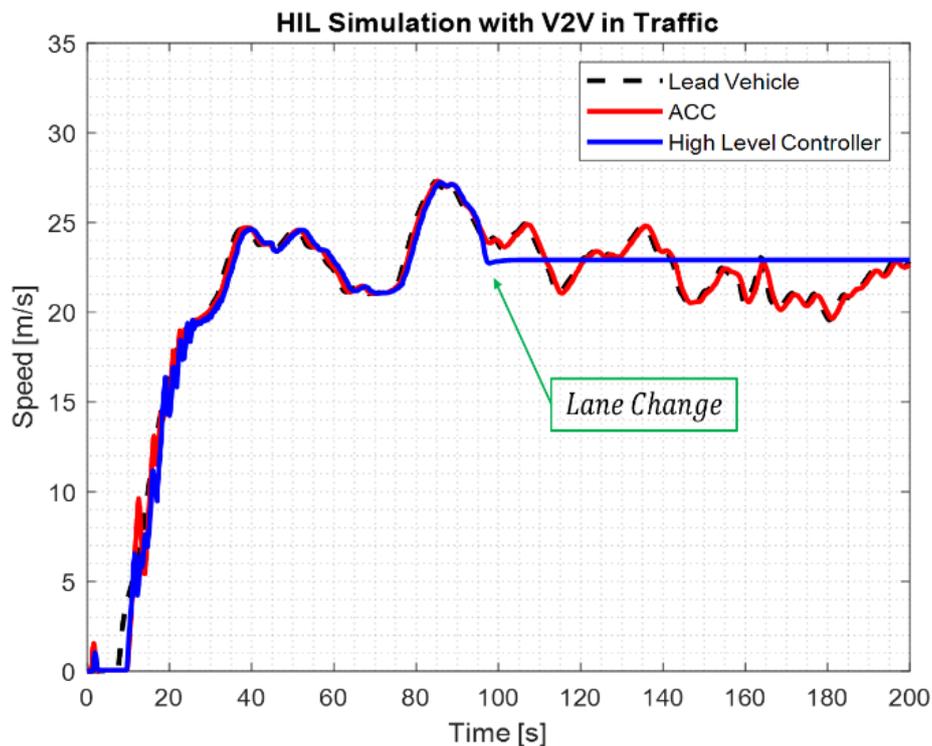

**Figure 13: HIL simulation with V2V in traffic results**

The results for these HIL simulations for car following are summarized in Table 5. Using the HL controller with CACC, Eco-CACC and Lane Change allowed the ego vehicle to have 4.36% more fuel savings compared to the case where the ego vehicle only had ACC for car following.

**Table 5: Fuel consumption for the HIL simulation with V2V in traffic**

| Model in HIL | Total Fuel Consumption [kg] | % Fuel Consumption Reduction wrt ACC |
|---|---|---|
| ACC | 0.156 | - |
| HL Controller | 0.149 | 4.36% |

## 4. Conclusions and Future Work

In this paper, 3 different car following models were studied. The ACC model aimed to keep a constant time gap between the ego and lead vehicles. CACC also aimed to keep a constant time gap, however, CACC also utilized the lead vehicle acceleration that is received through V2V.

Eco-CACC model perceived the lead vehicle acceleration that is received through V2V as a disturbance and aimed to attenuate it for car following while keeping a larger time gap compared to ACC and CACC.

Looking at the results, it is seen that CACC and Eco-CACC had superior performance and reduced the fuel consumption for the ego vehicle as compared to the ACC model. The HIL simulations were realistic and applicable to real-life implementation, since actual hardware experienced communication delay during V2V. The High Level (HL) controller utilized the lead vehicle driving behavior to determine with car following model was active, further improving the fuel economy.

For future work, the CACC and Eco-CACC models can be modeled in a Model Predictive Controller (MPC). This way, a cost function can be added to the Eco-CACC model to improve the fuel economy performance of the Eco-CACC further.

# 5. References

Aksun-Güvenç, B., Guvenc, L., 2002. The Limited Integrator Model Regulator and Its Use in Vehicle Steering Control. Turkish Journal of Engineering and Environmental Sciences 26, 473–482.

Aksun-Guvenc, B., Guvenc, L., 2002. Robust steer-by-wire control based on the model regulator, in: Proceedings of the International Conference on Control Applications. Presented at the Proceedings of the International Conference on Control Applications, pp. 435–440 vol.1. https://doi.org/10.1109/CCA.2002.1040225

Aksun-Guvenc, B., Guvenc, L., 2001. Robustness of disturbance observers in the presence of structured real parametric uncertainty, in: Proceedings of the 2001 American Control Conference. (Cat. No.01CH37148). Presented at the Proceedings of the 2001 American Control Conference. (Cat. No.01CH37148), pp. 4222–4227 vol.6. https://doi.org/10.1109/ACC.2001.945640

Aksun-Guvenc, B., Guvenc, L., Ozturk, E.S., Yigit, T., 2003. Model regulator based individual wheel braking control, in: Proceedings of 2003 IEEE Conference on Control Applications, 2003. CCA 2003. Presented at the Proceedings of 2003 IEEE Conference on Control Applications, 2003. CCA 2003., pp. 31–36 vol.1. https://doi.org/10.1109/CCA.2003.1223254

Aksun-Guvenc, B., Kural, E., 2006. Adaptive cruise control simulator: a low-cost, multiple-driver-in-the-loop simulator. IEEE Control Systems Magazine 26, 42–55. https://doi.org/10.1109/MCS.2006.1636309

Altan, O.D., Wu, G., Barth, M.J., Boriboonsomsin, K., Stark, J.A., 2017. GlidePath: Eco-Friendly Automated Approach and Departure at Signalized Intersections. IEEE Transactions on Intelligent Vehicles 2, 266–277. https://doi.org/10.1109/TIV.2017.2767289

Altay, I., Aksun Guvenc, B., Guvenc, L., 2014. Car Following with Adaptive Cruise Control Evaluated in İstanbul City Traffic Conditions. Presented at the 7th Automotive Technologies Congress – OTEKON 2014, Bursa, Turkey.

Asadi, B., Vahidi, A., 2011. Predictive Cruise Control: Utilizing Upcoming Traffic Signal Information for Improving Fuel Economy and Reducing Trip Time. IEEE Transactions on Control Systems Technology 19, 707–714. https://doi.org/10.1109/TCST.2010.2047860

Cantas, M.R., Fan, S., Kavas, O., Tamilarasan, S., Guvenc, L., Yoo, S., Lee, J.H., Lee, B., Ha, J., 2019a. Development of Virtual Fuel Economy Trend Evaluation Process, in:

WCX SAE World Congress Experience. SAE International. https://doi.org/10.4271/2019-01-0510

Cantas, M.R., Gelbal, S.Y., Guvenc, L., Guvenc, B.A., 2019b. Cooperative Adaptive Cruise Control Design and Implementation (SAE Technical Paper No. 2019- 01–0496). SAE International, Warrendale, PA. https://doi.org/10.4271/2019-01-0496

Demirel, B., Güvenç, L., 2010. Parameter Space Design of Repetitive Controllers for Satisfying a Robust Performance Requirement. IEEE Transactions on Automatic Control 55, 1893–1899. https://doi.org/10.1109/TAC.2010.2049280

Emirler, M.T., Güvenç, L., Güvenç, B.A., 2018. Design and Evaluation of Robust Cooperative Adaptive Cruise Control Systems in Parameter Space. Int.J Automot. Technol. 19, 359–367. https://doi.org/10.1007/s12239-018-0034-z

Emirler, M.T., Wang, H., Aksun Güvenç, B., Güvenç, L., 2016. Automated Robust Path Following Control Based on Calculation of Lateral Deviation and Yaw Angle Error. Presented at the ASME 2015 Dynamic Systems and Control Conference, American Society of Mechanical Engineers Digital Collection. https://doi.org/10.1115/DSCC2015-9856

Güvenç, L., Aksun Güvenç, B., Demirel, B., Emirler, M.T., 2017. Control of mechatronic systems, London. ed. IET.

Güvenç, L., Srinivasan, K., 1994. Friction compensation and evaluation for a force control application. Mechanical Systems and Signal Processing 8, 623–638. https://doi.org/10.1006/mssp.1994.1044

Güvenç, L., Uygan, I.M.C., Kahraman, K., Karaahmetoglu, R., Altay, I., Sentürk, M., Emirler, M.T., Hartavi Karci, A.E., Aksun Guvenc, B., Altug, E., Turan, M.C., Tas, Ö.S., Bozkurt, E., Ozguner, Ü., Redmill, K., Kurt, A., Efendioglu, B., 2012. Cooperative Adaptive Cruise Control Implementation of Team Mekar at the Grand Cooperative Driving Challenge. IEEE Transactions on Intelligent Transportation Systems 13, 1062–1074. https://doi.org/10.1109/TITS.2012.2204053

Kavas-Torris, O., Cantas, M.R., Gelbal, S.Y., Aksun-Guvenc, B., Guvenc, L., 2021. Fuel Economy Benefit Analysis of Pass-at-Green (PaG) V2I Application on Urban Routes with STOP Signs. International Journal of Vehicle Design (IJVD), Special Issue on Safety and Standards for Connected and Autonomous Vehicles 83, 258–279. http://dx.doi.org/10.1504/IJVD.2020.115058

Kianfar, R., Augusto, B., Ebadighajari, A., Hakeem, U., Nilsson, J., Raza, A., Tabar, R.S., Irukulapati, N.V., Englund, C., Falcone, P., Papanastasiou, S., Svensson, L., Wymeersch, H., 2012. Design and Experimental Validation of a Cooperative Driving System in the Grand Cooperative Driving Challenge. IEEE Transactions on Intelligent Transportation Systems 13, 994–1007. https://doi.org/10.1109/TITS.2012.2186513

Kural, E., Aksun Güvenç, B., 2010. Model Predictive Adaptive Cruise Control, in: 2010 IEEE International Conference on Systems, Man and Cybernetics. Presented at the 2010 IEEE International Conference on Systems, Man and Cybernetics, pp. 1455–1461. https://doi.org/10.1109/ICSMC.2010.5642478

Kural, E., Güvenç, B.A., 2015. Integrated Adaptive Cruise Control for Parallel Hybrid Vehicle Energy Management. IFAC-PapersOnLine, 4th IFAC Workshop on Engine and Powertrain Control, Simulation and Modeling E-COSM 2015 48, 313–319. https://doi.org/10.1016/j.ifacol.2015.10.045

Kural, E., Hacibekir, T., Aksun-Guvenc, B., 2020. State of the Art of Adaptive Cruise Control and Stop and Go Systems. arXiv:2012.12438 [cs].

Liu, H., Shladover, S.E., Lu, X.-Y., Kan, X. (David), 2020. Freeway vehicle fuel efficiency improvement via cooperative adaptive cruise control. Journal of Intelligent Transportation Systems 0, 1–13. https://doi.org/10.1080/15472450.2020.1720673

Ma, F., Wang, J., Yang, Y., Wu, L., Zhu, S., Gelbal, S.Y., Aksun-Guvenc, B., Guvenc, L., 2020. Stability Design for the Homogeneous Platoon with Communication Time Delay. Automot. Innov. 3, 101–110. https://doi.org/10.1007/s42154-020-00102-4

Ma, F., Wang, J., Yu, Y., Wu, L., Liu, Z., Aksun-Guvenc, B., Guvenc, L., 2021. Parameter-space-based robust control of event-triggered heterogene-ous platoon. IET Intelligent Transport Systems 15, 61–73. https://doi.org/10.1049/itr2.12004

Meier, J.-N., Kailas, A., Adla, R., Bitar, G., Moradi-Pari, E., Abuchaar, O., Ali, M., Abubakr, M., Deering, R., Ibrahim, U., Kelkar, P., Vijaya Kumar, V., Parikh, J., Rajab, S., Sakakida, M., Yamamoto, M., 2018. Implementation and evaluation of cooperative adaptive cruise control functionalities. IET Intelligent Transport Systems 12, 1110–1115. https://doi.org/10.1049/iet-its.2018.5175

Naus, G.J.L., Vugts, R.P.A., Ploeg, J., van de Molengraft, M.J.G., Steinbuch, M., 2010. String-Stable CACC Design and Experimental Validation: A Frequency-Domain Approach. IEEE Transactions on Vehicular Technology 59, 4268–4279. https://doi.org/10.1109/TVT.2010.2076320

Oncu, S., Karaman, S., Guvenc, L., Ersolmaz, S.S., Ozturk, E.S., Cetin, E., Sinal, M., 2007. Robust Yaw Stability Controller Design for a Light Commercial Vehicle Using a Hardware in the Loop Steering Test Rig, in: 2007 IEEE Intelligent Vehicles Symposium. Presented at the 2007 IEEE Intelligent Vehicles Symposium, pp. 852–859. https://doi.org/10.1109/IVS.2007.4290223

Orun, B., Necipoglu, S., Basdogan, C., Guvenc, L., 2009. State feedback control for adjusting the dynamic behavior of a piezoactuated bimorph atomic force microscopy probe. Review of Scientific Instruments 80, 063701. https://doi.org/10.1063/1.3142484

Wang, Z., Wu, G., Barth, M.J., 2018. A Review on Cooperative Adaptive Cruise Control (CACC) Systems: Architectures, Controls, and Applications, in: 2018 21st International Conference on Intelligent Transportation Systems (ITSC). Presented at the 2018 21st International Conference on Intelligent Transportation Systems (ITSC), pp. 2884–2891. https://doi.org/10.1109/ITSC.2018.8569947

Wei, Z., Hao, P., Barth, M.J., 2019. Developing an Adaptive Strategy for Connected Eco-Driving under Uncertain Traffic Condition, in: 2019 IEEE Intelligent Vehicles Symposium (IV). Presented at the 2019 IEEE Intelligent Vehicles Symposium (IV), pp. 2066–2071. https://doi.org/10.1109/IVS.2019.8813819

Xiao, L., Gao, F., 2010. A comprehensive review of the development of adaptive cruise control systems. Vehicle System Dynamics 48, 1167–1192. https://doi.org/10.1080/00423110903365910

Yang, Y., Ma, F., Wang, J., Zhu, S., Gelbal, S.Y., Kavas-Torris, O., Aksun-Guvenc, B., Guvenc, L., 2020. Cooperative ecological cruising using hierarchical control strategy with optimal sustainable performance for connected automated vehicles on varying road conditions. Journal of Cleaner Production 123056. https://doi.org/10.1016/j.jclepro.2020.123056